# TiO$_2$ with Super Narrow Bandgap Achieved in One-Step Single-Mode Magnetic Microwave Induced Plasma Treatment

Kunihiko Kato [a], Yunzi Xin [b], Takashi Shirai [a,b*]

[a] *Department of Life Science and Applied Chemistry, Graduate School of Engineering, Nagoya Institute of Technology, Gokiso, Showa-ku, Nagoya, Aichi 466-8555, Japan*
[b] *Advanced Ceramics Research Center, Nagoya Institute of Technology, Gokiso, Showa-ku, Nagoya, Aichi 466-8555 Japan*
*Corresponding authors: e-mail: shirai@nitech.ac.jp*

## Abstract

TiO$_2$ with super narrow bandgap (1.1 eV~) are successfully synthesized via a facile and novel one-step single-mode magnetic microwave induced plasma treatment. The selectively surface Ti$^{3+}$-doping on obtained TiO$_2$ as trapping centers which significantly restrain the photo-excited carrier recombination and achieve the enhancement of visible-light photocatalytic performance. In addition, the surface chemical composition of TiO$_x$ is precisely controlled in a wide region of $1.19 < x < 1.92$ during one-step reaction. A detailed XPS analysis reveals that the surface formed Ti$^{3+}$ shows highly thermal/chemical stability even through high-temperature treatment (~800 °C) in oxidative atmosphere and photocatalytic reaction.

## Keywords

Building up specific physical/chemical structure in material outermost surface is one of essential key point for design of advanced functional materials. Photocatalyst has attracted considerable attention as a material for environmental purification, whose activity is strongly affected by surface chemical state since photo-catalytic reactions arise on material outermost surface. In the case of titanium dioxide ($TiO_2$), which has been widely used as photocatalyst, much efforts paid for designing band structure by preparing composite structure [1,2] and doping anions or cations into $TiO_2$ lattice as one of widely known strategy to assembling visible-light $TiO_2$ photocatalyst [3-7]. However, especially these doped type $TiO_2$ possess large amount of carrier recombination sites, which cause to deteriorate the photocatalytic activity remarkably and suffer from thermal instability [8,9]. Recently, a reduced type $TiO_2$ with highly concentrated defects such $Ti^{3+}$ species and/or oxygen vacancies have attracted much interest in regard to its excellent visible light photocatalytic performance towards $CO_2$ reduction, $H_2$ production, and photo-degradation of hazardous organic material [10-12]. However, in the general synthesis methods as represented by liquid phase synthesis such sol-gel or hydro/solvothermal method [13,14], it is difficult to generate highly concentrated defects on surface selectively, the generated bulk defects cause recombination between photo-excited carriers, resulting in decrease of photocatalytic activity in comparison of surface defects [15]. Additional calcination process is also needed for crystallization of produced $TiO_2$ to enhance photocatalytic activity, this process can cause to an extinction of defects. In the case of heat treatment for reduction of $TiO_2$, sever experimental condition is necessary such utilization of explosive gas ($H_2$) as reducing gas or/and vacuum system for long time [16-18].

Here in the present research, we report a facile method for the synthesis of chemically/thermally stable deficient $TiO_2$ with tunable chemical composition towards highly efficient visible-light photocatalyst from commercially available micrometer size rutile-$TiO_2$ via one-step single-mode magnetic microwave (SMMW) irradiation induced plasma treatment The obtained $TiO_2$ exhibits sufficient wide-ranging light absorption from visible-light to near infrared region with super narrow bandgap. In addition, the obtained $TiO_2$ shows superior photocatalytic performance in photo-degradation of Rhodamine B (RhB) under visible light irradiation. Furthermore, the mechanism of surface defects introduced on $TiO_2$ by SMMW plasma is systemically discussed as based on the specific heat history of target material during the SMMW irradiation and detailed characterization via X-ray photoelectric spectroscopy (XPS).

Figure 1a shows the UV-vis absorption spectra of the surface modified $TiO_2$ at varied temperature. The obtained $TiO_2$ exhibit superior light absorption in wide region from UV to visible light wavelength (~800 nm) to compare with raw $TiO_2$. Also the obtained $TiO_2$ shows the obvious color changes from white to black after the short-time treatment as shown in Figure 1b. Generally, such a black coloration of $TiO_2$ can be observed in reduced type of $TiO_2$, which have highly concentrated defects as $Ti^{3+}$ and $Ti^{2+}$ species and/or oxygen vacancies [11,19,20]. It is known that electrons are more likely to be trapped at these surface defect sites on $TiO_2$, and it causes light absorption in specified wavelength as color center, especially visible light. Furthermore, the modified $TiO_2$ show narrower optical bandgap, according to the deduced Tauc-plot shown in Figure 1c. The values of optical bandgap are summarized as a function of treatment temperature in Figure 1d. In the case of the $TiO_2$ obtained via treatments under 700 ºC, it is in good agreement with the reduced type of $TiO_2$, where $Ti^{3+}$ and oxygen vacancy induce the localized levels at 0.75 and 1.18 eV below conduction band minimum (CBM), respectively [21-23]. However, MW-800 exhibited super narrow optical bandgap about 1.1 eV, which has been never observed in micrometer size $TiO_2$ to our knowledge.

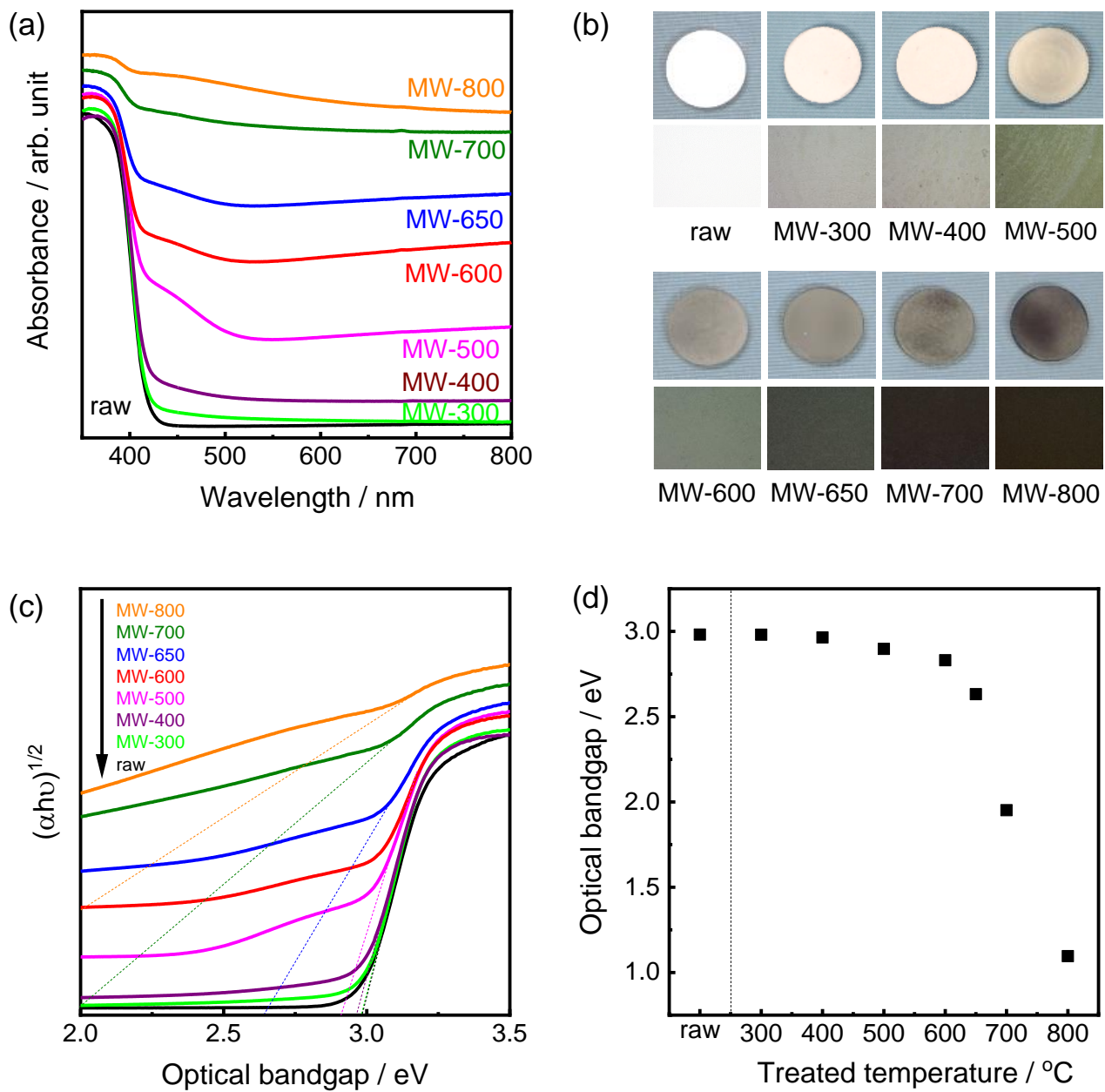

**Figure 1.** (a) Absorbance spectra in UV-vis, (b) Photographs (upside) and optical microscope images (downside), and (c) Tauc-plot of the obtained $TiO_2$. (d) Value of optical bandgap deduced from Tauc-plot.

The surface chemical state of the modified $TiO_2$ are characterized from XPS spectra of $Ti_{2p}$ orbital. As shown in Figure 2a and Figure S1, the spectra gradually broadened apparently after SMMW plasma treatment. These peaks are fitted based on Gaussian equation with three components at 458.8, 457.1 and 455.5 eV, which can be assigned as $Ti^{4+}$, $Ti^{3+}$ and $Ti^{2+}$, respectively [24,25]. Figure 2b displays that the atomic concentration of defects originated from unoccupied bonding state ($Ti^{3+}$ and $Ti^{2+}$) increase linearly as function of the temperature of target $TiO_2$ during SMMW induced plasma treatment. Additionally, the chemical composition of the modified $TiO_2$ are illustrated as Figure 2c. It can be demonstrated that the chemical composition can be precisely controlled as $TiO_x$ with value x in a wide range of 1.19 to 1.92.

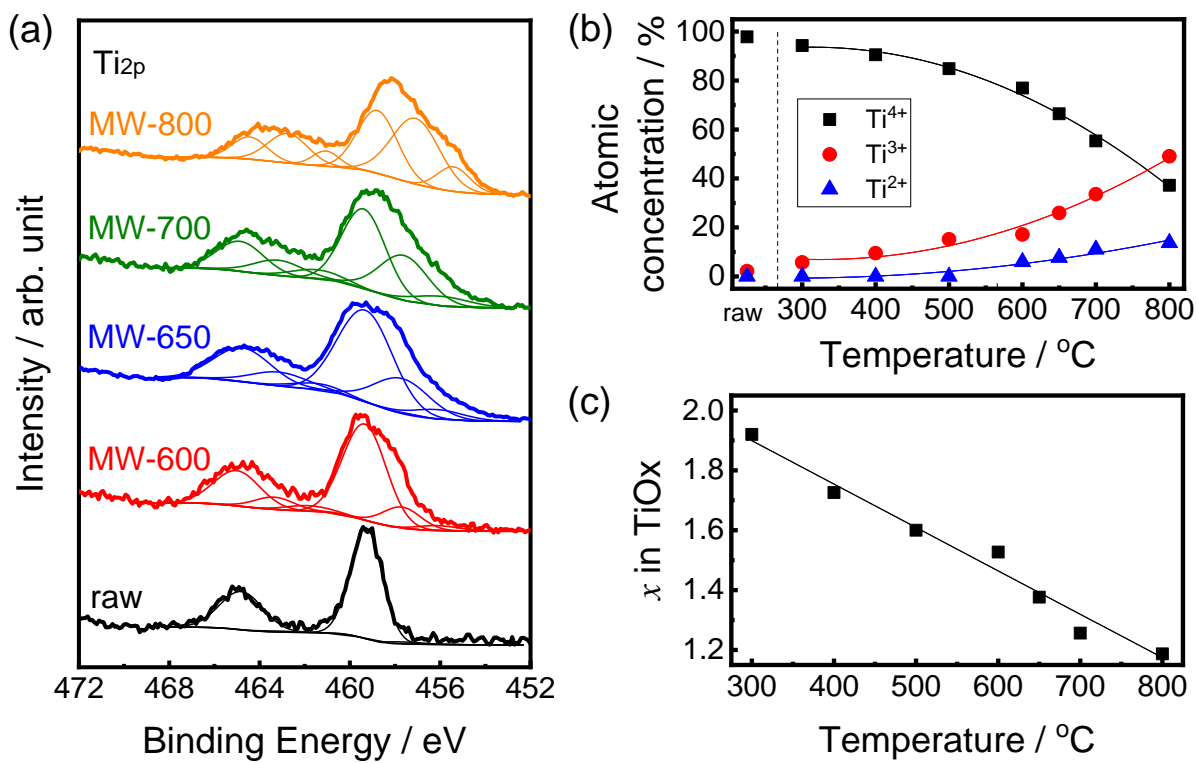

**Figure 2.** (a) The XPS spectra in $Ti_{2p}$ orbital of the obtained $TiO_2$ via He plasma induced by SMMW irradiation, (b) The atomic concentration of Ti charge species ($Ti^{4+}$, $Ti^{3+}$, and $Ti^{2+}$), (c) The chemical composition of the modified $TiO_2$.

Figure 3a shows the photocatalytic performance of the obtained $TiO_2$ in photo-degradation of Rhodamine B under visible light irradiation. As a result, a significant improvement of photocatalytic activity can be achieved after the treatment. PL spectra are measured to characterize the efficiency of charge recombination and separation between the photo-excited carriers on the surface of modified $TiO_2$ [26,27]. As shown in Figure 3b, the modified $TiO_2$ exhibited lower recombination rate than commercial $TiO_2$, since the PL intensity decreased significantly. It is known that the surface defect sites on $TiO_2$ derived from $Ti^{3+}$ species can work as electron trap centers. Thus, these sites lead to restrain the recombination between photo-excited carriers and increase the lifetime of excited electrons as well [28-30]. In addition, the defects can also improve the electric conductivity [31-33] and promote the electron transfer. Therefore, it could be expected that the surface modified $TiO_2$ via SMMW plasma induced treatment performs high photocatalytic efficiency with the existence of sufficiently separated photo-excited carriers even in micrometer size rutile-$TiO_2$.

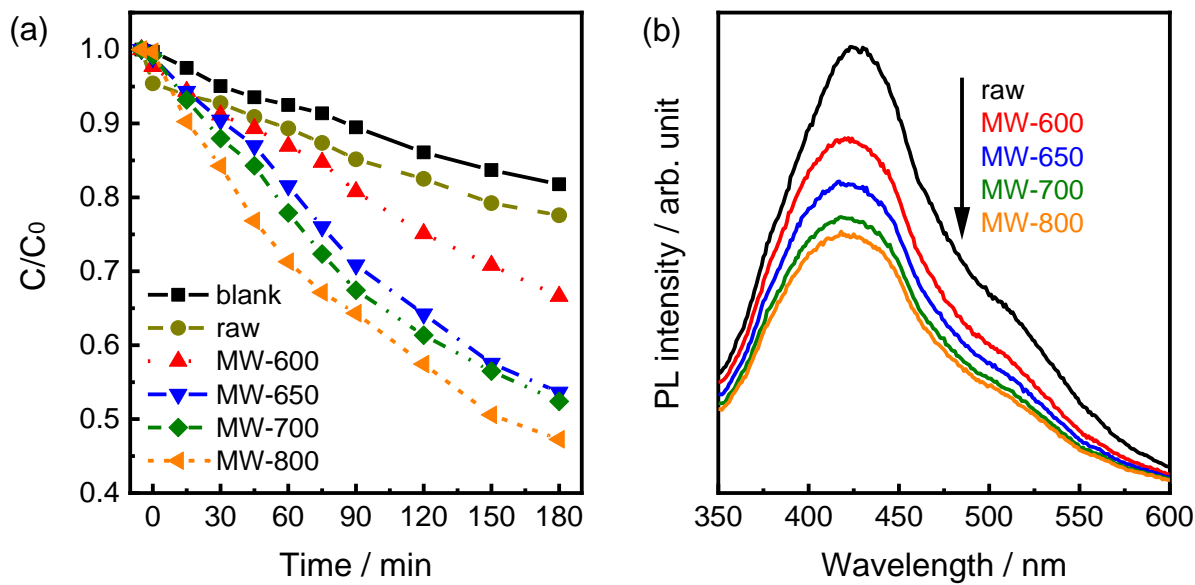

**Figure 3.** (a) Photocatalytic activity towards the degradation of Rhodamine B under irradiation of visible light for constant time ($C/C_0$ are the percentage photo-degradation of RhB), (b) PL spectra of the surface modified $TiO_2$.

The schematic figure for SMMW induced plasma treatment towards highly deficient $TiO_2$ is shown in Figure 4. A possible mechanism is expected as the following description. Firstly, a high energy plasma induced by SMMW field collide towards $TiO_2$ surface, which cause the dissociation of Ti-O bonding and generate defects such as $Ti^{3+}$ species and oxygen vacancies on $TiO_2$ surface by releasing oxygen atoms. Actually, this expected mechanism agrees with the results observed also in the case of $TiO_x$ obtained via a treatment using $N_2$ gas as a plasma resource instead of He gas. Figure S2a, b, and c shows the XPS spectra of $Ti_{2p}$, $O_{1s}$, and $N_{1s}$, respectively. The peaks, which are assigned to $Ti^{3+}$, $Ti^{2+}$ and $OH^-$, appear and gradually become dominant with increase of treatment temperature as shown in $Ti_{2p}$ and $O_{1s}$ spectra. In addition, the existence of N atom can be observed as interstitial N in $TiO_2$ lattice whose assigned peaks is located at 399 eV [34,35]. This result supports the above description for the mechanism that the SMMW induced plasma collided toward $TiO_2$ surface and O atoms is released. The atomic concentration and x value in $TiO_x$ are summarized in Figure S2d, e, and f. The highly concentrated N atoms (17.94 atom %) are also confirmed to be injected in the obtained $TiO_2$ after MW plasma treatment. Furthermore, the obtained $TiO_2$ exhibit highly visible light absorption via plasma treatment using $N_2$ plasma according to absorbance spectra of UV-vis shown in Figure S3a. Yellowish coloration derived from existence of N atoms into $TiO_2$ lattice also can be observed in N-doped $TiO_2$ as shown in Figure S3b. Interestingly in this process, although $TiO_2$ should not be heated in magnetic MW field, the temperature of target $TiO_2$ rise with increase of MW output according to the temperature history during MW irradiation as shown in Figure S4 in the both cases of inducing He and $N_2$ plasma, respectively. It is expected that plasma intensity depend on MW output, much heat would be generated by strong colliding of MW plasma towards $TiO_2$ surface. Surprisingly, highly concentrated defects remain stably on $TiO_2$ surface even though the surface temperature of target $TiO_2$ reached at high temperature (~800 °C) during SMMW plasma treatment. Actually, the large amount of lower titanium charge species ($Ti^{3+}$ and $Ti^{2+}$) are still remained in the sample after the photocatalytic characterization as shown in Figure S5. Generally, these surface defects existed on $TiO_2$ can be disappeared easily by heat treatment at more than 500 ºC [36,37]. Additionally, especially surface $Ti^{3+}$ species are highly unstable and can be easily disappeared via oxidation in air or water environment [12]. This stability of defected sites can be seen also in the case of N-doped $TiO_2$ synthesized via $N_2$ plasma treatment. Since $N_{int}$ atom is generally considered to be more thermodynamically unstable than a substitutional N ($N_{sub}$) due to lower activation energy, resulting in diffusing and disappearing easily in heat treatment at high temperature [38]. However, even although the synthesized $TiO_2$ underwent high temperature treatment around 800 ºC during SMMW assisted reaction, the doped $N_{int}$ atom still exist stably without transferring to $N_{sub}$. Therefore, the generated defect sites or injected atoms during SMMW plasma treatment exist stably on the surface modified $TiO_2$ than the generally known reduced or doped type $TiO_2$. In addition, in comparison of other reported methods using plasma induced by high frequency wave (RF and microwave) [21,39,40], the $TiO_2$ synthesized via SMMW plasma shows more highly concentrated defects for shorter treatment time without the utilization of reducing gas such $H_2$ in our method.

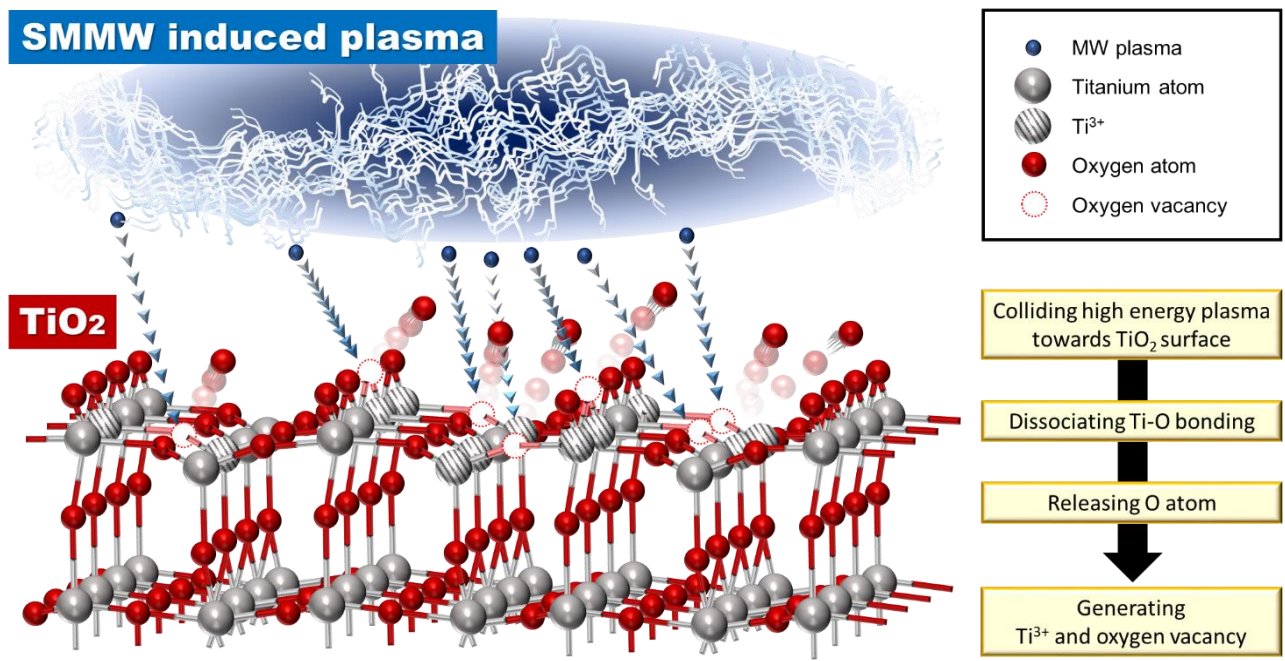

**Figure 4.** Schematic figure for the mechanism of SMMW induced plasma treatment toward $TiO_2$.

In conclusion, highly-stable surface deficient $TiO_2$ is successfully synthesized from commercial micrometer-sized rutile-$TiO_2$ through one-step SMMW induced plasma treatment. The highly $Ti^{3+}$ self-doped $TiO_2$ demonstrates sufficient wide-ranging light absorption (~800 nm) as well as super narrow bandgap (1.1 eV ~) accompanied with the black coloration, which has been never observed in micrometer size $TiO_2$. In addition, the chemical composition can be tunable precisely in the wide range of $1.19 < x < 1.92$ ($TiO_x$) by the condition of SMWW induced plasma treatment even without the utilization of reducing gas such $H_2$. Moreover, the obtained $TiO_2$ exhibit lower recombination rate between photo-excited electrons and holes, resulting in the superior photocatalytic performance for the photo-degradation of RhB under visible-light irradiation. Furthermore, the detailed XPS characterization clarified that $Ti^{3+}$ species formed on $TiO_2$ outermost surface can exist stably even through the severe environmental condition such high temperature thermal reaction (~800 ºC) and photocatalytic reaction accompanied with generation of strongly oxidative active species. The SMMW plasma assisted treatment would provide new routine for development of functional metal oxides with well-designed specific surface structure.